\documentclass[]{interact}

\usepackage{epstopdf}
\usepackage[caption=false]{subfig}

\usepackage[numbers,sort&compress]{natbib}
\bibpunct[, ]{[}{]}{,}{n}{,}{,}
\makeatletter
\def\NAT@def@citea{\def\@citea{\NAT@separator}}
\makeatother

\theoremstyle{plain}

\theoremstyle{definition}

\theoremstyle{remark}

\usepackage{url}
\usepackage{xcolor}

\begin{document}


\title{The Extremely Large Telescope}

\author{
\name{Paolo Padovani\thanks{CONTACT Paolo Padovani. Email: ppadovan@eso.org} and Michele Cirasuolo}
\affil{European Southern Observatory, Karl-Schwarzschild-Str. 2, D-85748 Garching bei M\"unchen, Germany}
}

\maketitle

\begin{abstract}
Extremely large telescopes (ELTs) are considered worldwide to be one of the 
highest priorities in ground-based astronomy. 
The European 
Southern Observatory (ESO) is 
developing an ELT that will have a 39 m main mirror and will be the largest 
visible and infrared light telescope in the world. 
The ELT will be equipped with a 
lineup of cutting-edge instruments, designed to cover a wide range of 
scientific possibilities. 
The leap forwards with the ELT can lead to a paradigm shift in our perception of the Universe, much as Galileo's telescope did 400 years ago. 
We illustrate here the various components of the ELT, including the dome and main structure,
the five mirrors, and the telescope systems. We then describe the ELT 
instrumentation and some of the astronomical topics it will address. We then 
conclude by examining the 
synergies with other astronomical facilities. 
\end{abstract}


\begin{keywords}
ELT; Extremely Large Telescopes; science; astronomy; technology; 
telescopes; solar system; exoplanets; stars; black holes; 
galaxies; cosmology; dark matter; fundamental physics. 
\end{keywords}

\section{Introduction}\label{sec:intro}

The past decade has brought astronomical discoveries that have excited people from 
all walks of life, from finding planets around Proxima Centauri, the nearest star to the 
Sun, to the first image of a black hole, namely that at the centre of
M 87, a giant elliptical galaxy in the Virgo constellation. In the next epoch of astronomy, the ELT\footnote{\url{https://elt.eso.org/}}, with the
other two extremely large telescopes, the 24.5 m Giant Magellan Telescope\footnote{\url{https://giantmagellan.org/}} and the Thirty Meter 
Telescope\footnote{\url{https://www.tmt.org}}, 
will tackle some of the biggest scientific challenges of our time. The ELT 
will track down Earth-like planets around other stars, and could become the first telescope 
to find evidence of life outside of our Solar System. It will also probe the furthest 
reaches of the cosmos, revealing the properties of the very earliest galaxies and the 
nature of the dark Universe. On top of this, astronomers are also planning for the unexpected 
-- new and unforeseeable questions that will surely arise, given the new capabilities of the ELT. 

The ELT programme was approved in 2012 and green light for construction at Cerro 
Armazones in Chile's Atacama Desert was given at the end of 2014. From building the immense telescope dome structure to casting of the mirrors, the work on this 
wonder of modern engineering has been made possible thanks to the spirit of collaboration. 
ESO has been working alongside a worldwide community and dozens of Europe's most cutting-edge 
companies to bring the ELT to ``technical first light'' later this decade 
(currently planned for the end of 2028).

The decision on the ELT site was based on an extensive comparative meteorological 
investigation, which lasted several years. The  selection process took several 
factors into account, with the ``astronomical quality'' of the atmosphere, that is, 
the number of clear nights, infrared (IR) properties (height, 
temperature, precipitable 
water vapour, etc.), seeing, atmospheric turbulence profile, mean coherence length 
and time, outer scale length, playing a crucial role. The 
selection also considered the operation and scientific synergy with other major 
facilities, such as the Very Large Telescope (VLT)  at ESO's 
Paranal Observatory and the Atacama Large Millimeter/submillimeter Array (ALMA). 
Finally, a significant role 
was played also by construction and logistical aspects, i.e. the impact of the 
location on construction cost and schedule, considerations on where the observatory 
staff would live, the closeness of power grids, water supply, 
political and seismic stability, and so on. 

Cerro Armazones' location in the dry Chilean desert at high altitude (3046 m)
made it ideal for astronomical observations, with over 320 clear nights per year and 
offering some of the darkest skies on Earth with virtually no light pollution. 
The altitude of the site above sea level 
does not pose logistical problems for operations, while meeting the science requirements 
for low precipitable water vapour and low operating temperatures. The median seeing is 
0.67 arcsecond at 500 nm with a median coherence time of 3.5 ms. The rainfall in one year 
is of the order of 100 mm, with a median relative humidity value of 15\%. Moreover,
it is located just 20 km away from Cerro Paranal, 
which hosts ESO's VLT and other telescopes. 

\section{Telescope Overview}\label{sec:telescope}

The telescope and its interior structure will be housed in the the giant ELT dome, 
which will provide protection from the extreme environment of Chile's Atacama Desert. 
The main structure of the telescope will hold its five mirrors and optics, including 
the enormous 39 m primary mirror. The ELT will employ sophisticated ``adaptive optics'' (AO) technologies 
to compensate for the turbulence of the Earth's atmosphere and to ensure its images 
are sharper than those of any other telescope. It will also have other components 
such as a prefocal station that works as the link between the telescope and its 
instruments. Finally, it will have a modern control system to allow the user to 
operate the telescope for science observations and maintenance activities. The ELT 
will be part of ESO's Paranal Observatory and will be operated from the same control 
room as other ESO telescopes such as the VLT.

\subsection{Dome and Main Structure}

\begin{figure}
\centering
\includegraphics[width=1.0\textwidth]{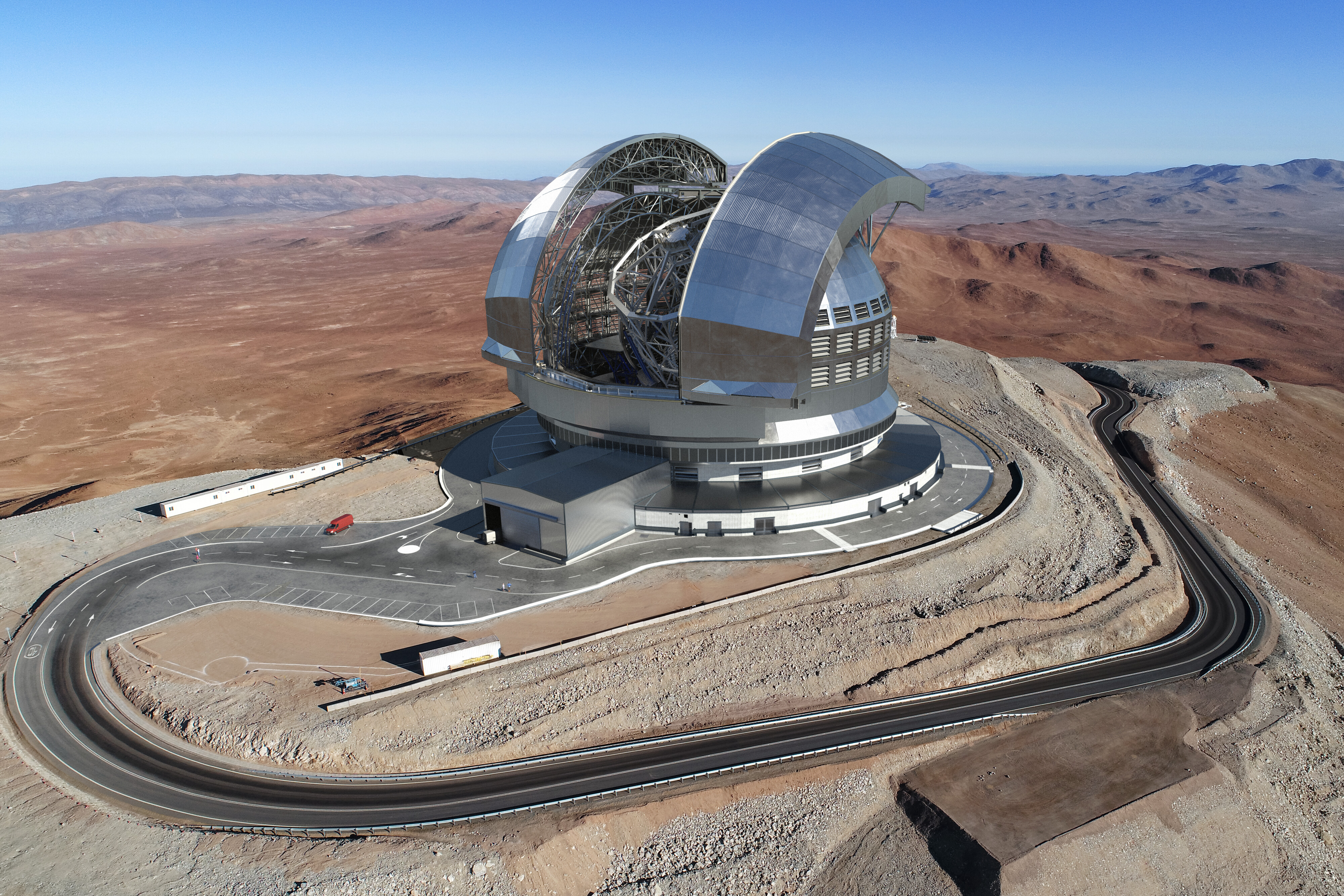}
\caption{Artistic rendering of how the ELT will look like on top of Cerro Armazones. {\it Image credit: ESO.}} 
\label{fig:dome}
\end{figure}

The sensitive and delicate components of the ELT 
require special protection. This is especially true in the 
rough desert environment, with potentially damaging sand and wind, and in a country 
that is prone to earthquakes. The ELT dome shelters the telescope in windy conditions up to 18 m/s
during observations and protects it and its delicate optics from the elements during 
the day. It consists of a lower fixed part, the concrete pier, a rotating upper part, 
and the enclosure, which is equipped with two motorised slit doors that open laterally 
during observations. The latter has a thermally insulated aluminium cladding, 
which enables thermal conditioning of the telescope chamber during 
the day and limits radiative cooling during the night. This rotating structure is also 
equipped with special cranes to handle operations during the day when the 
slits are closed. The pier of the dome is surrounded by an auxiliary building, with 
the instrument assembly area and the entrance hall facing south, downstream of the 
prevailing wind. The various electrical, thermal, and hydraulic plants used to operate 
the dome and 
the telescope are distributed in different rooms of the building. The remaining rooms 
are used for mirror segment storage, computer rooms, and to host a facility for mirror 
coating. 
The anti-seismic concept of the ELT telescope and dome is based on seismic isolators, 
which are placed below the telescope between the base layer of the foundation and the rock. 
In case of an earthquake, the telescope and the dome foundation can slide horizontally on 
those isolators by as much as 30 cm to compensate for the ground movements, minimising 
shocks to the telescope. The telescope has been designed to survive the most powerful
earthquakes ever recorded in Chile.

The telescope needs to be able to track about the 1$^{\circ}$ zenithal avoidance locus 
and preset to a 
new target within 5 minutes. This requires the enclosure to accelerate and move at angular speeds 
of up to 2$^{\circ}$/s (the linear speed at the outer diameter being approximately 5 km/h), which 
poses considerable challenges to the motorisation system and its braking devices, and demands the 
implementation of various safety provisions. The enclosure rotates independently from the telescope 
structure, and sufficient mechanical clearance is guaranteed to permit the complete movement of the 
telescope within the open or closed dome. The dome design enables observations from zenith down to 
20$^{\circ}$ from the horizon. To avoid vibration during the rotation of the enclosure, the 
rotating mechanism is, to the extent possible, structurally decoupled from the dome concrete base. 
In addition, the possible propagation of vibrations from the dome foundation to the telescope pier, 
via the seismic isolators and the rock substrate, has been studied to ensure a smooth and 
vibration-free movement and to guarantee the required precision of the telescope tracking during 
observation.

The rotating enclosure, once fully equipped and finished, will have a height and 
a diameter of 80 and 88 m respectively, a mass of around 6100 tons, and 
it will rotate on 36 stationary trolleys mounted above the pier at a height of 11 m from the 
ground (Figure \ref{fig:dome}). The structure itself consists of a round girder with a special track at its bottom that 
rests on the wheels of the trolleys. The primary structure of the enclosure is completed by three 
structural arches resting on the girder, one at each side of the slit and one at the back. 
The enclosure structure, which will be bolted together on site, is closed by a number of secondary 
beams which will permit the assembly of the insulated aluminium cladding. A complex series of 
accesses inside the structure and the slit doors allows engineers to reach all mechanisms of the 
doors, ventilation louvers, and installed equipment.
The enclosure's slit doors will move on 
three rails each, one at the round girder and two at the top. The doors provide an aperture of 
41 m when open, and their system motors have sufficient redundancy to ensure the doors can 
be closed when needed, in all conditions. In addition, the doors will be equipped with 
latching mechanisms to achieve structural continuity and with special inflatable seals 
to guarantee environmental tightness when closed.

\subsection{Mirrors}

\begin{figure}
\centering
\includegraphics[width=0.95\textwidth]{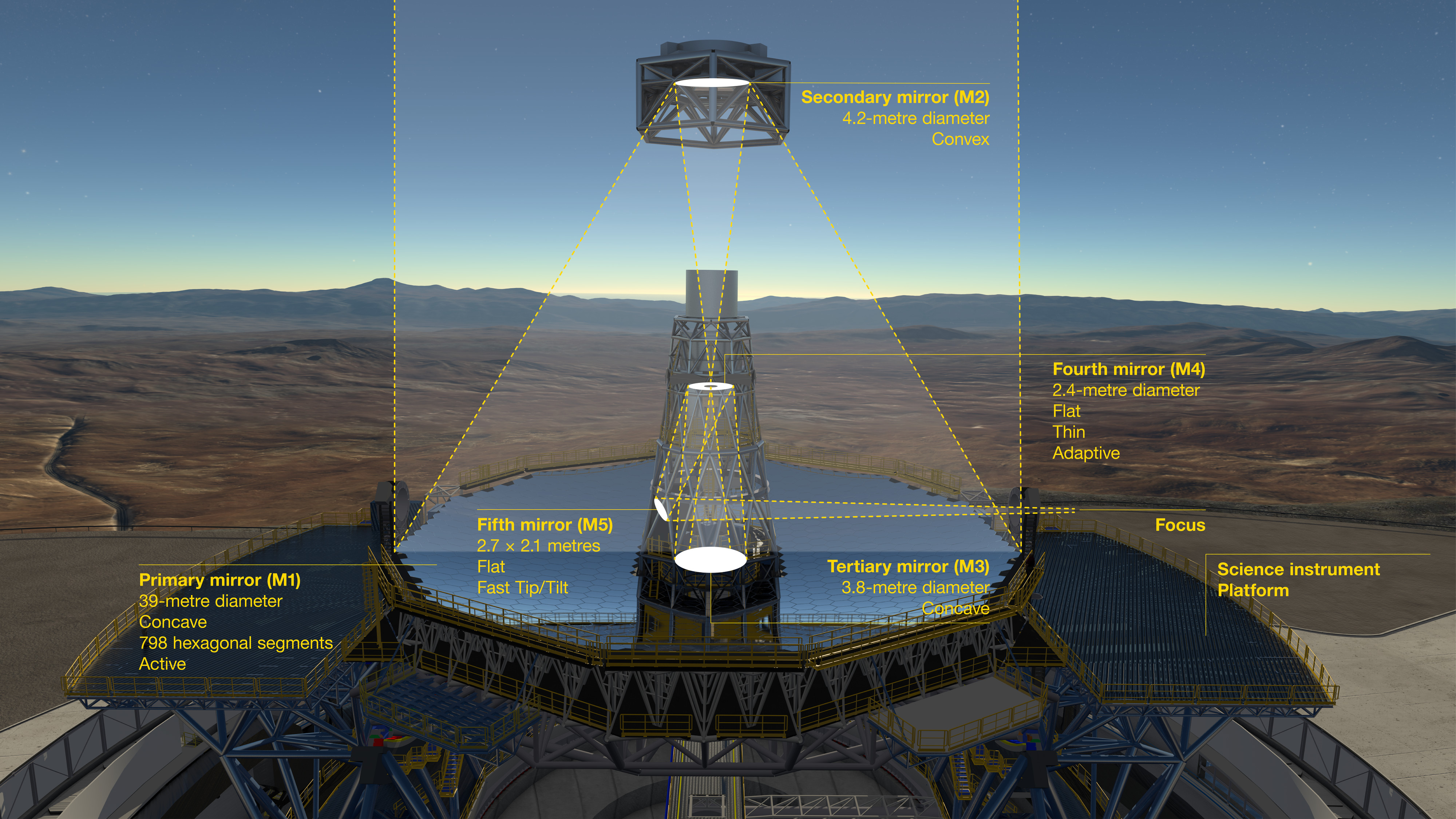}
\caption{The novel 5-mirror optical system of the ELT. Before reaching the science instruments the light is first reflected from the telescope's giant concave 39 m segmented primary mirror (M1), it then bounces off two further 4-metre-class mirrors, one convex (M2) and one concave (M3). The final two mirrors (M4 and M5) form a built-in AO system to allow extremely sharp images to be formed at the final focal plane. The two Nasmyth platforms to the left and to the right will host the science instruments and the pre-focal stations (Section \ref{sec:pfs}). {\it Image credit: ESO.}}
\label{fig:light_path}
\end{figure}

The ELT will have a pioneering optical design with five mirrors, all having different 
shapes, sizes, and roles but working together seamlessly. The primary, M1, is the 
most spectacular: a giant 39 m concave 
mirror that will collect light from the night sky and reflect it to the secondary mirror. The 
convex M2, the largest secondary mirror ever employed on a telescope, will hang above M1 and will 
reflect light back down to M3, which in turn will relay it to an adaptive flat mirror, M4, above 
it. This will adjust its shape a thousand times a second to correct for distortions 
caused by atmospheric turbulence, before sending the light to M5, a flat tiltable mirror that will 
stabilise the image and send it to the ELT instruments on one of the two Nasmyth platforms.
Figure \ref{fig:light_path} shows the optical path of the telescope with the five mirrors.

This design provides an unvignetted field of view (FoV) with a diameter of 10 arcminutes on the 
sky, or about 80 square arcminutes (i.e. $\sim 1/9$ of the area of the full moon). Thanks to the 
combined activation of M4 and M5, it will have the capability to correct for atmospheric 
turbulence as well as the vibration of the telescope structure induced by its movement and the 
wind. This is crucial to allow the ELT to reach its diffraction limit, which is $\sim$ 8 
milli-arcseconds (mas) in the J-band ($\lambda \sim 1.2 \umu$m) and $\sim$ 14 mas in the K-band 
($\lambda \sim 2\umu$m), thereby providing images 15 times sharper than the Hubble Space Telescope 
(HST).

\subsubsection{M1}\label{sec:M1}

\begin{figure}
\centering
\vspace{-1.8cm}
\includegraphics[width=0.95\textwidth]{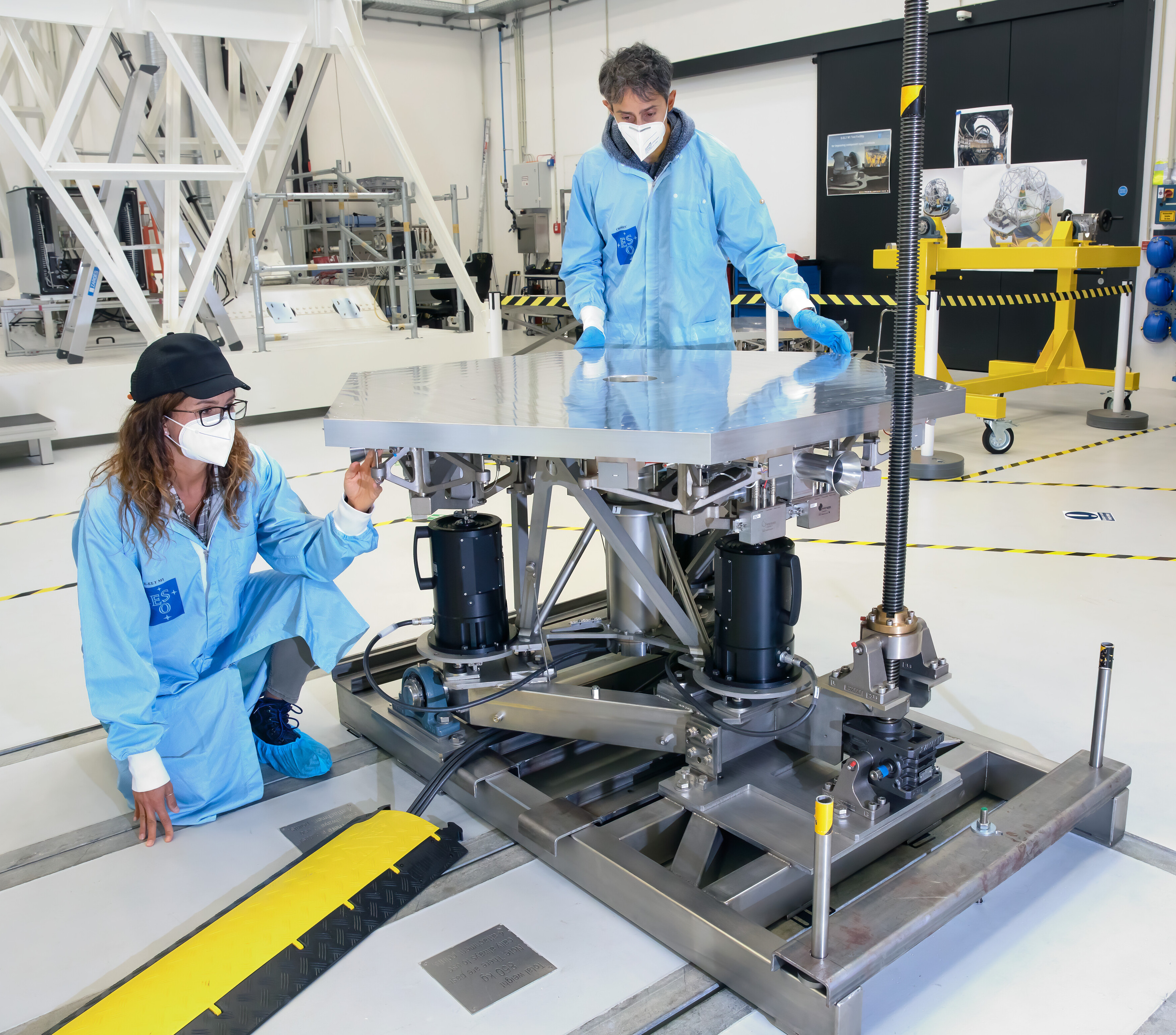}
\caption{A dummy M1 segment assembly in the ESO Integration Lab in Garching, Germany, with position actuators (in black), whiffletree (above
them), and two test edge sensors (Figure \ref{fig:edge}) just below the segment. {\it Image credit: ESO.}} 
\label{fig:whiffle}
\end{figure}

\begin{figure}
\centering
\includegraphics{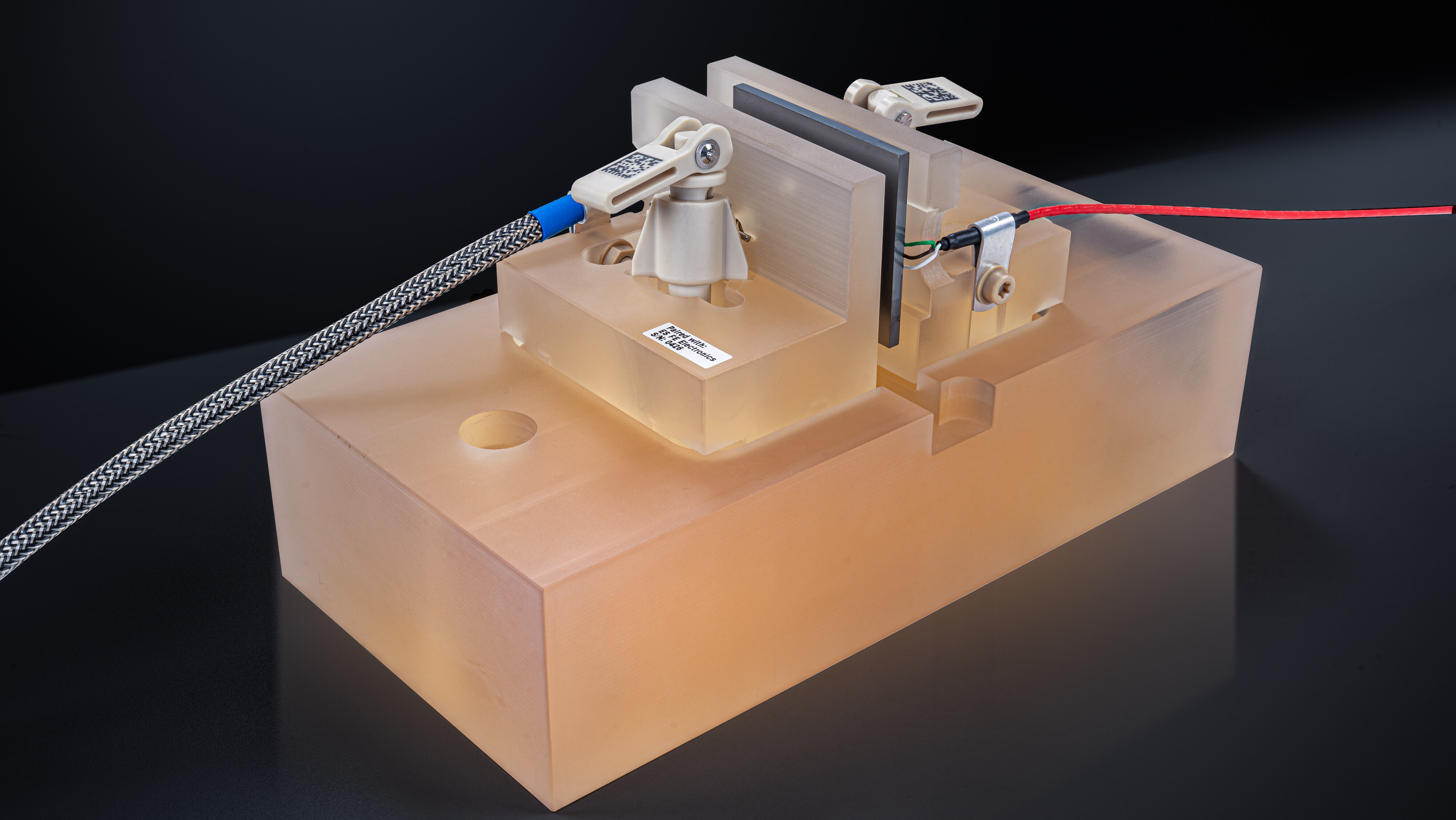}
\caption{The M1 edge sensors, which are designed and manufactured by the FAMES consortium (comprising Fogale Nanotech in France and Micro-Epsilon in Germany), are the most precise edge sensors ever designed for a telescope. Two on each side of each segment, the sensors can detect when the segments move out of position, even if only by a millionth of a millimetre. This photo shows an edge sensor test model. {\it Image credit: FAMES (Fogale Nanotech, Micro-Epsilon).}}
\label{fig:edge}
\end{figure}

The primary mirror, M1, at 39.3 m in diameter will be the largest 
optical/IR mirror ever built (Figure \ref{fig:light_path}). 
The mirror is segmented, being made of 798 quasi-hexagonal mirrors, 
each of which is about 1.45 m in size (corner-to-corner), only 50 mm thick and weighs 250 
kg. The full M1 has a six-fold symmetry, as there are six identical sectors of 133 segments each. 
In each sector all 133 segments are different from one another in shape and optical prescription; 
in other words there are 133 different segment types. In order to facilitate recoating there will 
be a seventh sector with 133 segments, i.e. one for each segment type. This adds up to a grand 
total of 931 segments. To achieve the required scientific performance, M1 needs to be maintained 
in position and be phased to an accuracy of tens of nm, i.e. 10,000 times thinner than 
a human hair, across its entire 39 m diameter. This is extremely challenging, as the full 
structure will be moving constantly during an observation, and will be affected by wind and 
thermal changes. To achieve this, each segment, made of the low-expansion ceramic material 
Zerodur$^\copyright$\footnote{This special material is not sensitive to 
thermal fluctuations thanks to its very low thermal expansion coefficient. This means that the 
form and the shape of the mirrors will not change significantly with temperature during 
observations. It is also extremely resistant, can be polished to the required 
finishing level, and has been used in telescope mirrors for decades.} (from SCHOTT), is 
supported on a 27-point (three 9 points) whiffletree, which is a mechanism to evenly 
distribute the support across the back of the segment using 27 points of contact across its 
surface (Figure \ref{fig:whiffle}). Each whiffletree includes a primary tripod and three secondary tripods (inner and outer secondary tripods), connected by flexures in a frictionless design. The axial support carries the loads between the segment and the moving frame along the segment local normal z axis.
The load can be adjusted via warping harnesses so as to 
slightly change the shape of the mirror to compensate for optical aberrations induced 
by gravity and thermal effects. Moreover, each segment assembly can be moved in height 
and tip/tilt relative to the structure by using three positioning actuators (PACTs). 
These three actuators can move independently with an accuracy of 2 nm with a maximum 
excursion (or stroke) of 10 mm to reposition the segment to its desired position, 
thereby compensating for the deflections of the underlying telescope structure
and maintain the perfect co-alignment of all 
the segments, effectively creating a giant monolithic mirror. 
In addition, the actuators need to provide the resolution and accuracy necessary for phasing the primary mirror and holding it against perturbations. These latter are dominated by the effect of wind across the front surface of the primary mirror and by possible vibrations arising from machinery either directly under the segments or transmitted through the structure to the primary segments. 
The co-alignment is done through two 
``edge sensors'' (Figure \ref{fig:edge}) on each side of each hexagonal segment that constantly measure piston, 
gap and shear with respect to the adjacent segment to nanometre accuracy and provide the 
necessary information to the control system to activate the PACTs, allowing the segments 
to work together to form a perfect imaging system. 
The edge sensors are non-contact devices which measure relative segment positions at the border between adjacent segment. They are made of two major components: 1. a sensing head, consisting of two parts, called by the generic terms ``transmitter'' and ``receiver''; 2. a control electronic, including all the necessary cabling, providing the piston, gap and shear measurements, as well as diagnostic data. The ``transmitter'' and the ``receiver'' are located on the adjacent edges of two neighbouring segments and are facing each other. There are six ``transmitters'' and six ``receivers'' per segment, except for the segments located at the inner and outer contours of the M1 mirror. All in all, M1 features  
798 segments, almost 2500 PACTs and about 9000 edge sensors (4500 pairs), 
not including the seventh sector and the spares. More details on the M1 mirror can be found in \citep{M1}. 

\subsubsection{M2 and M3}

The ELT's M2 mirror, with its 4.25 m diameter, will be the largest secondary mirror ever used on a 
telescope, and the largest convex mirror ever produced. For comparison, the secondary mirrors 
on the 8 m VLTs are just over 1 metre in diameter. M2 is larger than 
the primary mirror of ESO's VISTA telescope and indeed the primary mirrors
of many other telescopes that are operating today. There is also the added challenge that M2 
will hang upside-down over the 39 m M1 mirror, about 60 m above the ground, held in mid-air 
by its support structure (called the M2 cell) and anchored to the telescope main structure. 
The M3 mirror is similarly large and complex, with its 4 m diameter. The mirrors alone 
weigh more than 3 tonnes each but with the cell and structure the overall weight of each 
assembly is about 12 tonnes. Both M2 and M3 are produced by the German company SCHOTT and, as
is the case for M1, are also made of ZERODUR$^\copyright$. 

The M2 mirror is a convex 4.25 m F/1.1 thin meniscus, about 100 mm thick, with an 800 mm 
central hole. Its optical surface shape is very aspheric, with a departure from a sphere 
that is close to 2 mm. The size, convexity, aperture ratio and asphericity make this mirror 
extremely difficult to polish and test. The M3 mirror is a concave 4.0 m F/2.6 thin meniscus, 
about 100 mm thick, with a 30 mm central hole. Its optical surface shape is mildly aspheric, 
with a departure from a sphere of only about 30 $\umu$m. Besides its 4 m size, the M3 mirror 
is easier to manufacture and test compared to the M2, and the required M3 mirror production 
and metrology processes are more common.
More details on the M2 and M3 mirrors can be found in \citep{M2-M3}. 

\subsubsection{M4}\label{sec:M4}

\begin{figure}
\centering
\vspace{-1.8cm}
\includegraphics[width=0.95\textwidth]{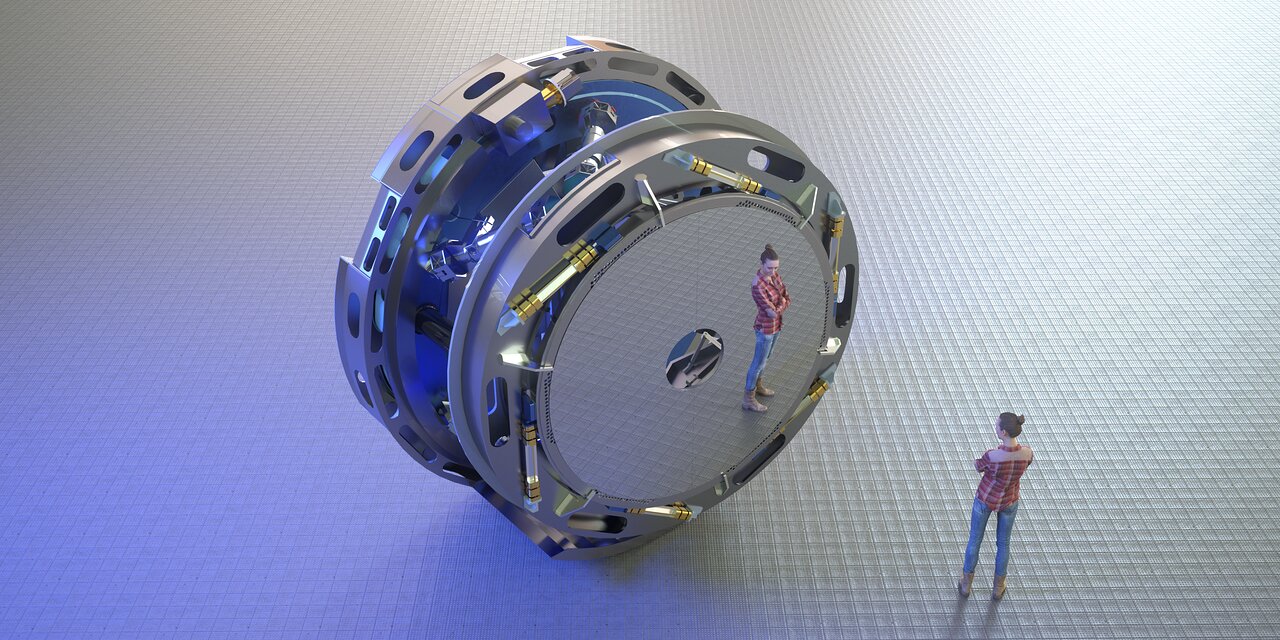}
\caption{A rendering of M4, the main adaptive mirror of the ELT. {\it Image credit: ESO.}}
\label{fig:M4}
\end{figure}

M4 (Figure \ref{fig:M4}) is the main adaptive mirror of the telescope, meaning that its surface can be deformed to 
correct for atmospheric turbulence, as well as for the fast vibration of the telescope 
structure induced by its motion and the wind, and is one of the most challenging and 
exciting ELT components.
More than 5,000 actuators will be used to 
change the shape of the mirror up to 1,000 times per second. In combination with the M5 mirror, 
M4 forms the core of the AO of the ELT (Section \ref{sec:AO}). With a diameter of 2.4 m, 
M4 will be the largest deformable
mirror ever built. By comparison, current adaptive mirrors are just over 1 m in diameter
(e.g., the 1.1 m M2 adaptive secondary on the VLT UT4 telescope). 

M4 consists of six ultra-thin segments, or petals, made of 
Zerodur$^\copyright$, like the M1 segments (Section \ref{sec:M1}). 
Because the M4 petals are so thin and need to be deformed to incredible accuracy, 
they require a very stable structure to support them: a reference body with attached 
magnets that support the mirror and adjust its shape. 
Initial studies have demonstrated that to achieve this highly demanding level of performance 
only specialist materials can be used for the mirror, either silicon carbide (SiC) or 
ultra-low-expansion glass machined in a special way to make it extremely lightweight. 
The French company Safran Reosc has developed a design for the M4 mirror using 
Boostec\textregistered~SiC. This material is well known for its high stiffness 
(stiffer than steel, carbon fibre or beryllium) and low density, properties that 
make the mirror very lightweight. SiC has been used for many space telescopes and 
the Herschel Space Observatory primary mirror is a good example of its 
technological feasibility.

Adaptive mirror technology was translated into an industrial astronomical product more 
than two decades ago by the Italian companies Microgate s.r.l and ADS, internationally 
known under the consortium name of AdOptica. Many 8 m telescopes now have a metre-scale 
adaptive mirror. The same technology is now being adapted to serve the ELT to 
produce a mirror with an area five times larger. The M4 mirror uses the same principle 
as a loudspeaker: the mirror is made of a very thin shell levitating 100 $\umu$m away 
from its reference surface  
and acting like a membrane, which deforms under the effect of about 5,000 voice coil actuators. 
These are a type of direct drive linear motor and the name ``voice coil'' comes from one of its  
first historical applications, vibrating the paper cone of a loudspeaker, consisting of a 
permanent magnetic field assembly and a coil assembly. The current flowing through the coil 
assembly interacts with the permanent magnetic field and generates a force that can be 
reversed by changing the polarity of the current. Depending on the current injected into the 
coil the mirror can be pushed or pulled up to a distance of 90 $\umu$m from its mean position. 
With the help of a very fast and precise set of capacitive sensors and amplifiers that are 
co-located with the voice coil actuators, the mirror's position is measured 70,000 times per 
second to an accuracy of a few tens of nm with the actuators being driven up to 1,000 times 
per second. More details on the M4 mirror can be found in \citep{M4}. 

\subsubsection{M5}\label{sec:M5}

M5 is the ``field stabilisation'' unit of the telescope. This means 
that the mirror is moving in a rigid way (tip-tilting) to steer the image and correct for 
vibrations of the telescope structure induced by its motion and by the wind, as well as 
some of the atmospheric turbulence. This is achieved by the M5 cell which is composed of 
three piezo actuators. It supports and moves the M5 mirror up to 10 times per second.
M5 is a flat, elliptical mirror with a diameter of 2.2 m on the minor axis and 2.7 m 
on the major axis. The M5 mirror assembly needs to be very 
stiff and at the same time very light (less than 500 kg in total) to allow its cell to move 
it fast enough whilst remaining flat when moving.
The French company Safran Reosc is manufacturing the M5 mirror assembly  
using Boostec\textregistered~SiC, as for the M4 mirror, while the Spanish company 
SENER Aeroespacial is responsible for the M5 cell.
More details on the M5 mirror can be found in \citep{M5}. 

\subsection{Telescope Systems}

\subsubsection{AO}\label{sec:AO}

Turbulence in the Earth's atmosphere causes the stars to twinkle in a way that 
blurs the finest details of the cosmos. Observing directly 
from space can avoid this atmospheric blurring effect, but the high cost of building,
launching, and 
operating 
space telescopes compared to using ground-based facilities limits the size and scope of 
the telescopes we can place off-Earth. Astronomers have turned to a method called adaptive optics. 
Sophisticated, deformable mirrors controlled by computers can correct in real-time for the 
distortion caused by the turbulence of the Earth's atmosphere, making the images obtained almost as 
sharp as (or, in the case of the ELT, sharper than) those taken in space. AO allows 
the corrected optical system to observe finer details of much fainter astronomical objects than is 
otherwise possible from the ground. AO requires a fairly bright reference star that is close 
to the object under study. This reference star is used to measure the blurring caused by the local 
atmosphere so that the deformable mirror can correct for it. Since suitable stars are not available 
everywhere in the night sky, astronomers can create artificial stars instead by shining a powerful 
laser beam into the Earth's upper atmosphere. Thanks to these laser guide stars, of which the ELT 
will have up to eight, almost the entire sky can now be observed with adaptive optics. 

There are various flavours of AO. Some of them are: 1. the classical Single Conjugated Adaptive Optics (SCAO) mode, where the turbulence corrected for is that in a cylinder projection of the telescope aperture towards a guide star. SCAO benefits from high performance, but across a small field of view; 2. the Laser Tomography Adaptive Optics (LTAO) mode, which is a technique similar to executing a computerized tomography scan of the atmosphere, as it allows the telescope to reconstruct the turbulence above it layer by layer. It uses laser guide stars instead of a natural one and 
implements several off-axis wavefront sensors; 3. the Multi-Conjugate Adaptive Optics (MCAO) mode, which  
uses multiple off-axis wavefront sensors to measure the turbulence in several directions. These measurements are fed to a tomographic reconstructor (using both multiple lasers and stars) that estimates the 3D turbulence, which is in turn projected onto the deformable planes to generate the control commands. This mode offers medium-to-high performance across a medium FoV.

From the largest adaptive mirror ever built to advanced control systems, the ELT will have some of 
the most sophisticated technologies ever employed on a telescope to correct for the blurring 
effects of the Earth's atmosphere. M4 and M5 are at the heart of adaptive optics corrections on ESO's ELT
(see Sects. \ref{sec:M4} and \ref{sec:M5}).

\subsubsection{Prefocal Stations}\label{sec:pfs}

\begin{figure}
\centering
\includegraphics[width=0.97\textwidth]{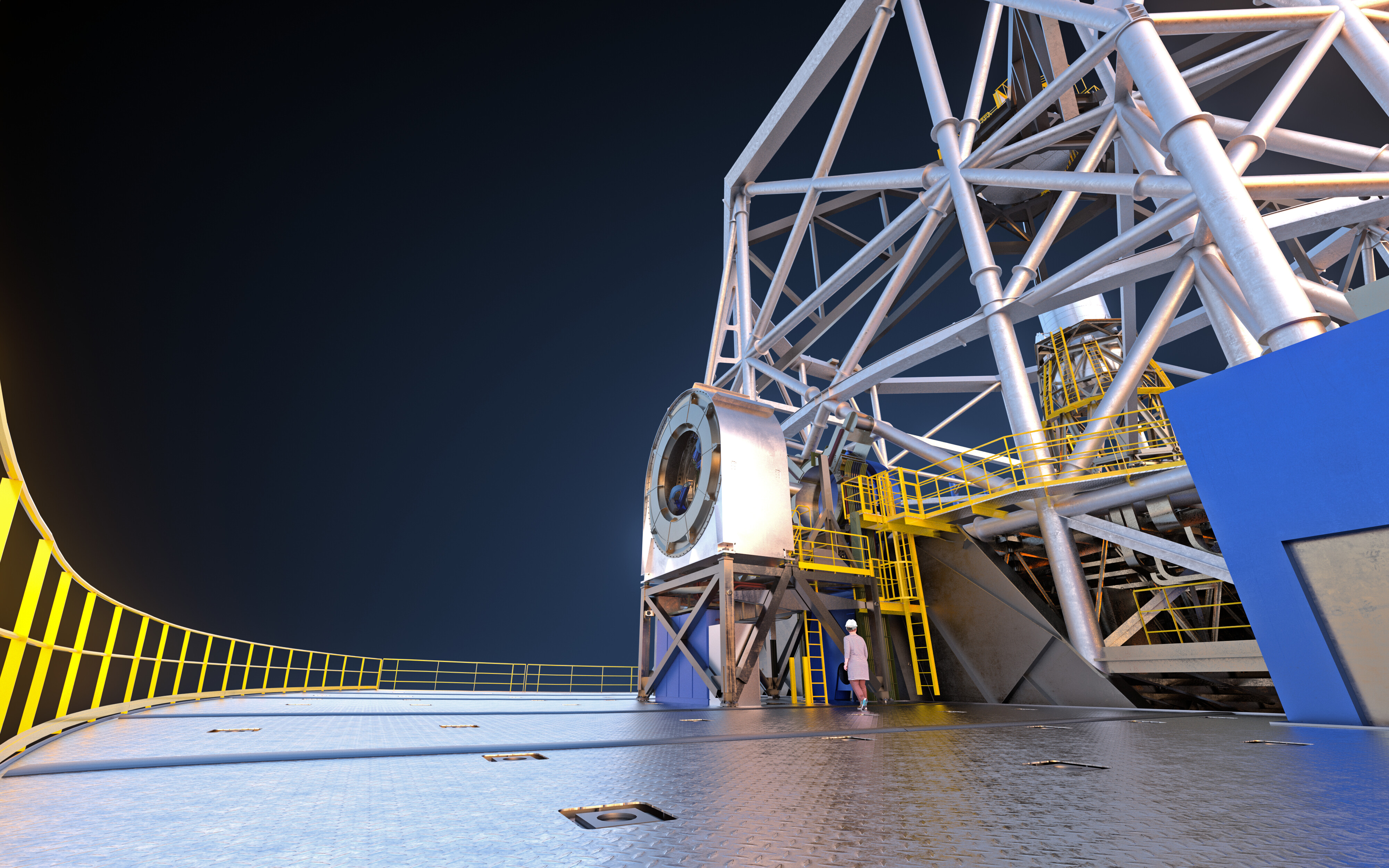}\hspace{5pt}
\caption{Computer rendering of the first ELT prefocal station (PFS-A) on top of its supporting platform. {\it Image credit: ESO/L. Cal\c{c}ada, IDOM.}}
\label{fig:PFS}
\end{figure}

The ELT requires not only sophisticated mirrors and advanced science instruments, 
but also a tight link between these two components. This link takes the form of two 
massive structures, the prefocal stations (PFS-A and PFS-B), which stand over 12~m 
high and sit on the platforms, called Nasmyth platforms, on either side of 
the ELT's giant tube structure (Figs. \ref{fig:PFS} and \ref{fig:light_path}). The name comes from the fact that the prefocal 
stations are the last components in the telescope's light path, right before the 
light comes to a focus. One of the most important functions of a prefocal station 
is to distribute light to the appropriate science instrument, which will then 
capture and analyse it. The instruments themselves are also on the two Nasmyth 
platforms, and the light is distributed to them via flat mirrors on the prefocal 
stations. Another main function of the PFS is sensing star light to control the 
telescope mirror alignment when observing. The PFS observes stars close to the 
target, known as guide stars, to check 
that the ELT's mirrors are aligned and that the telescope is correctly pointing at the target. 
The PFS also hosts a phasing and diagnostic station (PDS), which is equipped with special sensors 
used for testing the telescope before and during operation. As the name indicates, the PDS is used 
for general diagnostics, and to ensure that the giant primary mirror of the ELT (composed of 798 
segments) maintains an ideal shape and acts as one giant mirror, a process called phasing.


\subsubsection{Controls}

The many advanced mechanical and optical components, or subsystems, of the ELT, are operated by means 
of a control system. This complex system integrates the various components of the telescope and works 
as the user interface to the ELT. It provides coordinated, safe operation of the subsystems as a 
single system to perform science observations and support engineering and maintenance activities. 
The telescope control systems are also key in helping the ELT obtain sharp images. The complex 
primary mirror control system monitors and corrects the positions of its 798 segments, which are 
distorted by the effects of gravity and temperature, so it continues acting as one mirror (Section 
\ref{sec:M1}). The M4 (Section \ref{sec:M4}) and M5 (Section \ref{sec:M5}) mirror control systems 
are important for compensating for distortions in the images caused by 
atmospheric turbulence and wind, an important part of the ELT's adaptive optics.

Overall, the control system manages around 15,000 actuators and over 25,000 sensors 
distributed throughout the telescope and dome. The communication infrastructure allows 
the majority of the computing units to be hosted in the computer room in the auxiliary building of 
the dome. This reduces vibration and heat pollution in the telescope area and provides a controlled 
data centre environment for the hardware. 
The control system itself is a software-intensive system of systems, which presents a complex
challenge in terms of algorithms and computational performance,  
organisation of the overall system, and its behaviour and interactions.

\section{Instruments}\label{sec:instruments}

\begin{figure}
\centering
\includegraphics[width=1.0\textwidth]{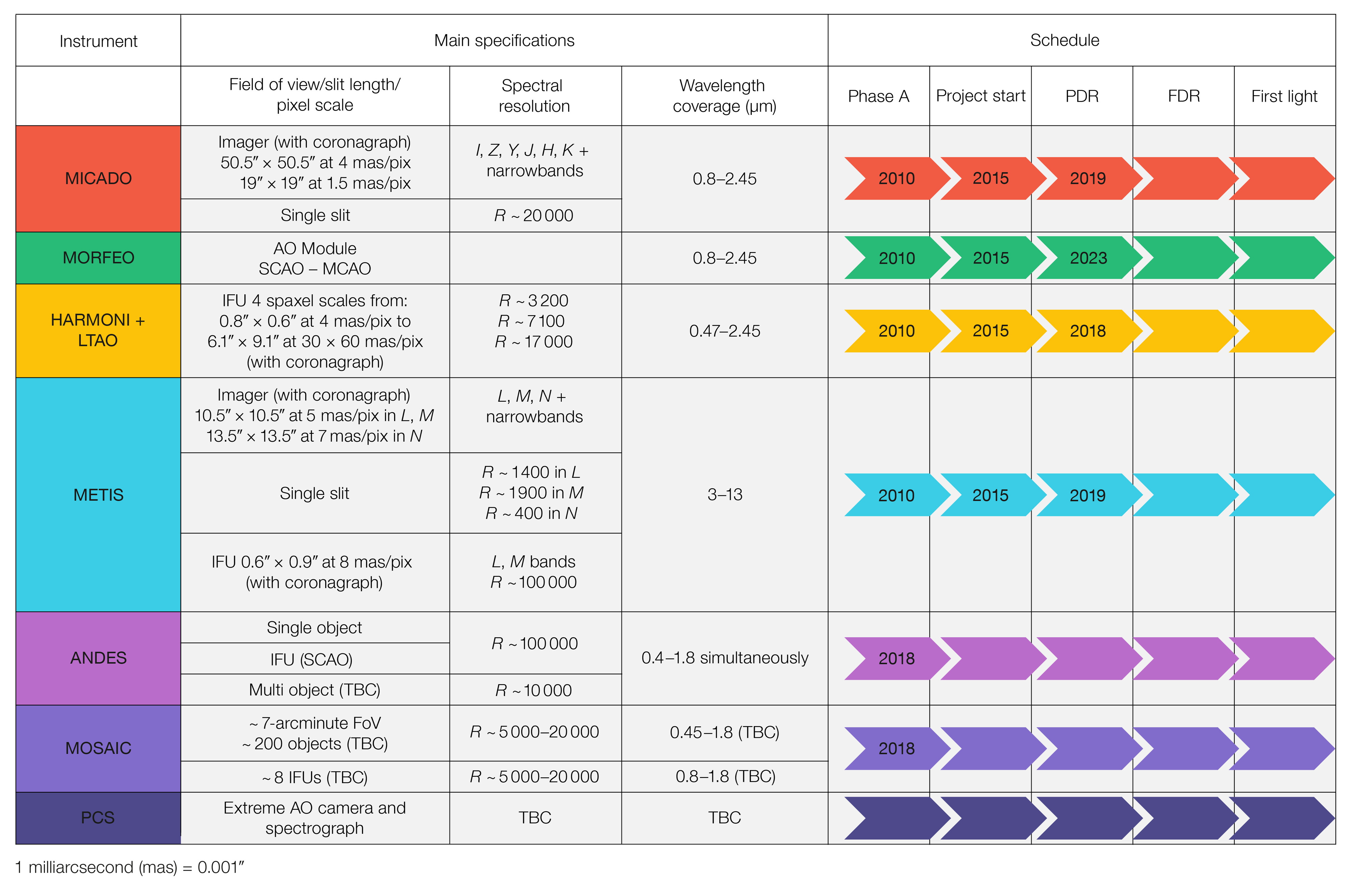}
\caption{A summary of the main features of the ELT instruments. PDR stands for
Preliminary Design Review while FDR stands for Final Design Review. {\it Image credit: ESO.}} 
\label{fig:inst}
\end{figure}

The suite of instruments planned for the ELT includes cameras, to capture images, and 
spectrographs, which disperse light into its component colours. 
Four of the instruments, HARMONI, MICADO, MORFEO, and METIS, the first generation, 
will start to operate at or shortly 
after ELT technical first light while an additional two, ANDES and MOSAIC, will 
start operations at a later stage. Throughout the telescope's lifetime, other 
instruments will be installed on the telescope to study the Universe in ever more 
detail, one of them being the Planetary Camera and Spectrograph (PCS: 
\citep{Kasper2021}). This will 
be dedicated to detecting and characterising nearby exoplanets with sizes 
from sub-Neptune to Earth-size in the neighbourhood of the Sun through a 
combination of eXtreme Adaptive Optics (XAO), coronagraphy, and spectroscopy. 
Figure \ref{fig:inst} summarizes the main features of ELT instruments. 

\subsection{HARMONI}

The High Angular Resolution Monolithic Optical and Near-infrared Integral field 
spectrograph (HARMONI: \cite{Thatte2021}) is the visible and near-infrared (NIR), 
adaptive-optics-assisted, integral field spectrograph for the ELT. It will have 
both an SCAO mode (using a single bright natural 
guide star) and an LTAO mode (using multiple 
laser guide stars: Section \ref{sec:AO}), to correct for atmospheric turbulence, providing near 
diffraction-limited hyper-spectral imaging. 
With its 3D spectroscopic capabilities and a variety of spatial and spectral settings, 
HARMONI will be a very powerful and versatile first-generation instrument.
HARMONI will use an Integral Field Unit (IFU) with an image slicer to divide a single 
contiguous FoV into many spatial pixels (called spaxels). The signal from each 
spaxel will be fed into a spectrograph that generates a spectrum for each one of them. 
Four spaxel scales (from 4 mas to 60 mas),  
resulting in different FoVs, will be available, with smaller spaxels corresponding to a 
smaller FoV but a higher spatial resolution. A range of spectral resolution settings 
(3000, 7000, 18000) will also be available in all spatial scales covering wavelengths 
between 0.47 and 2.45 $\umu$m, though not simultaneously. 

\subsection{MICADO}\label{sec:MICADO}

The Multi-AO Imaging Camera for Deep Observations (MICADO: \cite{Davies2021})
will make the most of the giant ELT's full resolution potential. It will 
take images at an unprecedented depth, thanks to the 
combination of the ELT's large primary mirror and the correction
that will be provided by the MORFEO (Section \ref{sec:MORFEO}) 
AO system. The instrument will go far beyond the 
capabilities of the best observatories that we have today, including HST, and 
its sensitivity will be comparable to that of the James Webb Space Telescope (JWST) but 
with six times the resolution. 

The design of MICADO was driven by a desire for high sensitivity, resolution, 
astrometric accuracy, and wide wavelength coverage spectroscopy. Offering almost an 
arcminute-squared FoV with 4 mas pixels to sample fully the diffraction 
limit of the 
ELT, MICADO takes advantage of the wide-field correction and uniform point spread 
function offered by the MCAO module, MORFEO. A second, 
finer, plate scale (1.5 mas) will be used to help MICADO reach challenging requirements 
on astrometric precision (50 $\umu$mas). In addition to a large number 
of broad and narrow band filters for imaging (up to 30) covering the 0.8-2.45 $\umu$m 
wavelength range, MICADO also offers a wide-band spectroscopic mode. In single-slit 
spectroscopy, MICADO covers this wavelength range with a 
resolving power of 20,000. In common with the other early ELT 
instruments, it will have a coronographic capability to block starlight so that dimmer 
objects close to these stars, exoplanets, for example, can be seen more easily. A 
second adaptive-optics mode, SCAO, will be provided by 
MICADO and MORFEO. This mode is particularly suited to observations with the 
coronagraph, as it will provide excellent image quality over a smaller FoV. 

\subsection{MORFEO}\label{sec:MORFEO}
The Multiconjugate adaptive Optics Relay For ELT Observations (MORFEO, formerly known as MAORY: \cite{Ciliegi2021}) will help compensate for the distortion of light caused by turbulence in the Earth's atmosphere which makes astronomical images blurry. MORFEO will not make observations itself; rather, it will enable other instruments, such as MICADO (Section \ref{sec:MICADO}) in the first instance, to take exceptional images. MORFEO will use both M4 and two other deformable mirrors and other state-of-the-art systems to correct for different layers of turbulence high above the ELT. To measure atmospheric turbulence, MORFEO will observe three natural guide stars around the scientific FoV of the camera (MICADO) but it will also use six laser guide stars to reach the demanding image quality required for MICADO's ambitious science goals. The distortion of light by atmospheric turbulence will be measured by wavefront sensors using newly developed detectors capable of reading images hundreds of times per second with low noise. MORFEO will use six large mirrors including the two extra post-focal deformable mirrors to deliver light from the focal plane of the telescope to the focal plane of MICADO. By tilting one of those mirrors, it will be possible to steer the light beam towards MICADO or a second client instrument. 

\subsection{METIS}

The Mid-infrared ELT Imager and Spectrograph (METIS: \cite{Brandl2021}) 
will provide the ELT with a unique window to the mid-infrared (MIR: 
$3 - 13~\umu$m) band. Its SCAO system will enable high contrast imaging and IFU 
spectroscopy at the diffraction limit of the ELT. 
The instrument consists of two separate units, one for the imager and another one 
for the spectrograph. METIS will provide: 1. imaging in the L/M ($3 - 5~\umu$m) band,
together with medium resolution ($R \sim 1,000$) slit spectroscopy as well as 
coronagraphy for high contrast imaging; 2. imaging in the N ($7.5 - 13.5~\umu$m) band,  
with low resolution ($R \sim$ a few hundred) slit 
spectroscopy as well as coronagraphy for high contrast imaging; 3. high resolution 
($R \sim 100,000$) IFU spectroscopy in the L/M band, including a mode with extended 
instantaneous wavelength coverage (assisted by a coronagraph). 

\subsection{ANDES}

The ArmazoNes high Dispersion Echelle Spectrograph (ANDES, formerly known as HIRES: 
\cite{Marconi2021}) baseline design is that of a modular fibre-fed cross dispersed 
echelle spectrograph with three ultra-stable spectral arms, visual and NIR, providing a 
simultaneous spectral range of $0.4 - 1.8~\umu$m at a spectral resolving power 
$R \sim 100,000$ for a single object. ANDES will also include an IFU mode fed by an 
SCAO module to correct for the blurring effect of turbulence in the atmosphere. 
ANDES will separate light from the ELT mirrors into the wavelength channels using 
dichroic filters. Each wavelength channel interfaces with several fibre bundles that 
feed the corresponding spectrograph module (visual and NIR). 

\subsection{MOSAIC}

The Multi-Object Spectrograph (MOSAIC: \cite{Hammer2021}) is a versatile multi-object
spectrograph that will use the widest possible FoV provided by the ELT. It will
have three operating modes covering observations in the visible and IR bands 
for more than one hundred sources simultaneously. Namely: 1. a high multiplex 
mode in the visible (HMM-VIS), with simultaneous integrated-light observations 
of almost 200 objects, with the possibility of using both medium and high spectral 
resolving power ($R \sim 5,000$ and 20,000); 2. a high multiplex mode in the NIR 
(HMM-NIR), with simultaneous integrated-light observations of 80 objects (goal: 100) 
each of them with dedicated sky fibres to subtract the strong sky background present 
in the NIR. This mode will cover simultaneously the $0.8 - 1.8~\umu$m wavelength range 
with the possibility to observe in medium and high spectral resolving power; 3. 
a high-definition mode (HDM), with simultaneous observations of eight IFUs (goal: 
10 IFUs) deployed within a $\sim 40$ arcminute patrol field. Each IFU will cover 
a 2.5 arcsecond hexagon with $\sim 200$ mas spaxels, sharing the same spectrograph 
as for the HMM-NIR, with the same wavelength coverage and spectral resolving power.

\section{Science}\label{sec:science}

The ELT will tackle the biggest scientific challenges of our time, and will 
aim for a number of notable firsts, including tracking down Earth-like planets 
around other stars in the habitable zones where life could exist, which is one of 
the Holy Grails of modern observational astronomy. It will also make fundamental 
contributions to cosmology by probing the nature of dark matter and dark energy. 
Other key science areas include the study of stars in our Galaxy and beyond, 
black holes, evolution of distant galaxies, up to the very 
first galaxies in the so-called ``Dark Ages'', the earliest epoch of the Universe, which started 
380,000 years after the Big Bang, and watching the Universe expand in real time. 
On top of this astronomers are also planning 
for the unexpected: new and unforeseeable questions will surely arise, given the 
capabilities of the ELT.

\subsection{Solar System}

Although the most detailed information regarding objects in our solar system has 
generally come from in situ (or close-by) interplanetary missions (to, e.g., Venus, 
Mars, Jupiter, Saturn, Pluto and Charon, comet 67P/Churyumov-Gerasimenko), 
ground-based and space observatories have complemented these studies with new
discoveries, increasing their scientific return. The ELT will be no exception, as it 
will peer through planetary atmospheres, study volcanic activity and watery plumes on the 
moons of Jupiter and Saturn, and probe the origin of asteroids and comets. By taking 
repeated high-resolution images and spectra of planets and moons with evolving 
surfaces and atmospheres, the telescope will be capable of assembling a unique atlas of 
hundreds of Solar System objects. 

As the ELT will have a much larger sensitivity and resolution than the current generation 
of large telescopes, it will be particularly useful for studying the faintest objects in 
the Solar System, including those in the outer System, Neptune, Uranus and outer 
asteroids and comets, many of which have not been explored in detail with space missions. 
Furthermore, objects in the Asteroid Belt between Mars and Jupiter and the Kuiper Belt 
beyond Neptune 
provide a window into the properties of the protoplanetary disc from which Earth and the 
other planets formed 4.5 billion years ago. Using the ELT to explore these objects, we will 
gain a better insight into the creation and evolution of the Solar System.

\subsection{Exoplanets}

\begin{figure}
\vspace{0.2cm}
\centering
\vspace{-3.62cm}
\includegraphics[width=1.1\textwidth]{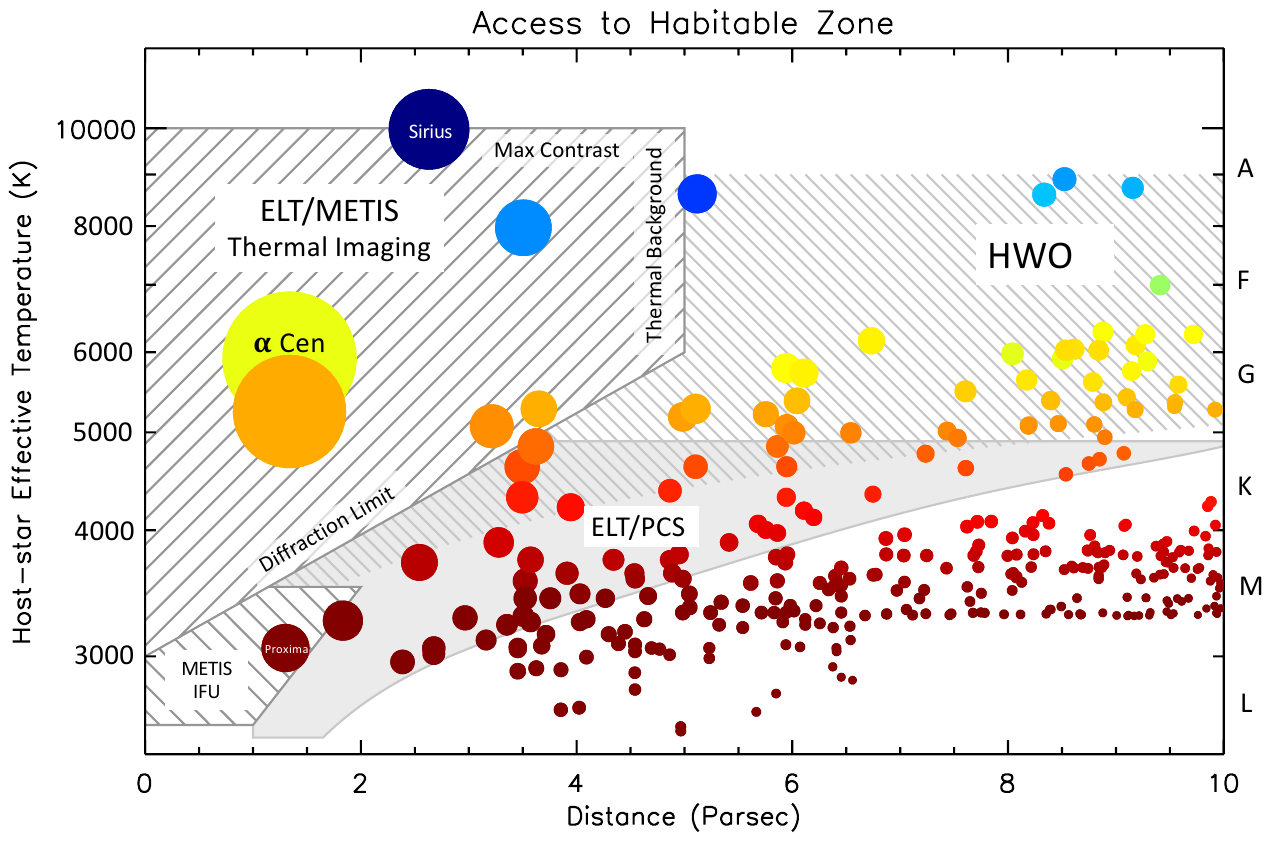}
\vspace{-2cm}
\caption{Effective temperature of the host star (or brown dwarf) estimated via the Stefan-Boltzman law vs. the distance to the system in parsecs. The thermal background limit of METIS for reasonable integration times sets the largest distance to which Earth-like planets can be detected in the liquid-water zone. As the equilibrium temperature of a planet increases with stellar luminosity, the habitable zone as bounded by the diffraction limit can be surveyed to greater distances around more luminous stars. The maximum contrast limits a search for Earth-like planets in the liquid-water zone to stars with $M < 2-3~M_{\odot}$, where 
$M_{\odot}$ is one solar mass. The METIS IFU can offer improved contrast at high spectral resolution enabling detection of unique planets like those around Proxima Centauri. The second generation ELT/PCS instrument will be more sensitive and will study liquid-water zone planets in scattered light. It will have  some overlap with a NASA future Great Observatory recently dubbed the Habitable Worlds Observatory (HWO). The colour of the points codes temperature from blue (hot) to red (cool). The size of the circles are related to the solid angle of the source, i.e. its surface area divided by its distance. {\it Image credit: Ignas Snellen and the METIS Science Team.}}
\label{fig:METIS}
\end{figure}

The first exoplanet orbiting a solar-type star (51 Pegasi) was discovered in 1995 by 
Michel Mayor and Didier Queloz, who received the Nobel Prize in Physics in 2019 for 
this breakthrough\footnote{\url{https://www.eso.org/public/announcements/ann19049/}}. 
Since then, almost 6,000 planets (at the time of this writing) have been found orbiting 
stars other than our Sun, from super Jupiters to Earth and Mars-size planets.  

Most exoplanets are detected indirectly by the radial velocity method, which detects 
planets by the wobble they produce on their parent star as they orbit it. Another 
method used by astronomers is the transit technique, which detects the drop in 
brightness of the host star when a planet passes in front of it. However, these 
detections are indirect and very limited in the information they can provide about 
the planet itself. In addition, very few direct observations of planets have been made.  

The ELT will revolutionise the study of planets outside our Solar System, as it will 
allow us to obtain direct images of some of these systems. Thanks to the very large 
collecting power of the ELT, we will also be able to detect and characterise 
the fingerprints of transiting planets' atmospheres as they pass in front of their parent star. 
Moreover, the ELT will allow us to observe and characterise 
exoplanets in habitable zones, i.e. in the regions where the 
distance from the central star is such that liquid water can exist on the planet's surface (Figure \ref{fig:METIS}). 
It could even become the first telescope to find evidence of extraterrestrial life on these planets.

\subsection{Stars}

\begin{figure}
\vspace{0.2cm}
\centering
\vspace{-3.62cm}
\includegraphics[width=0.925\textwidth]{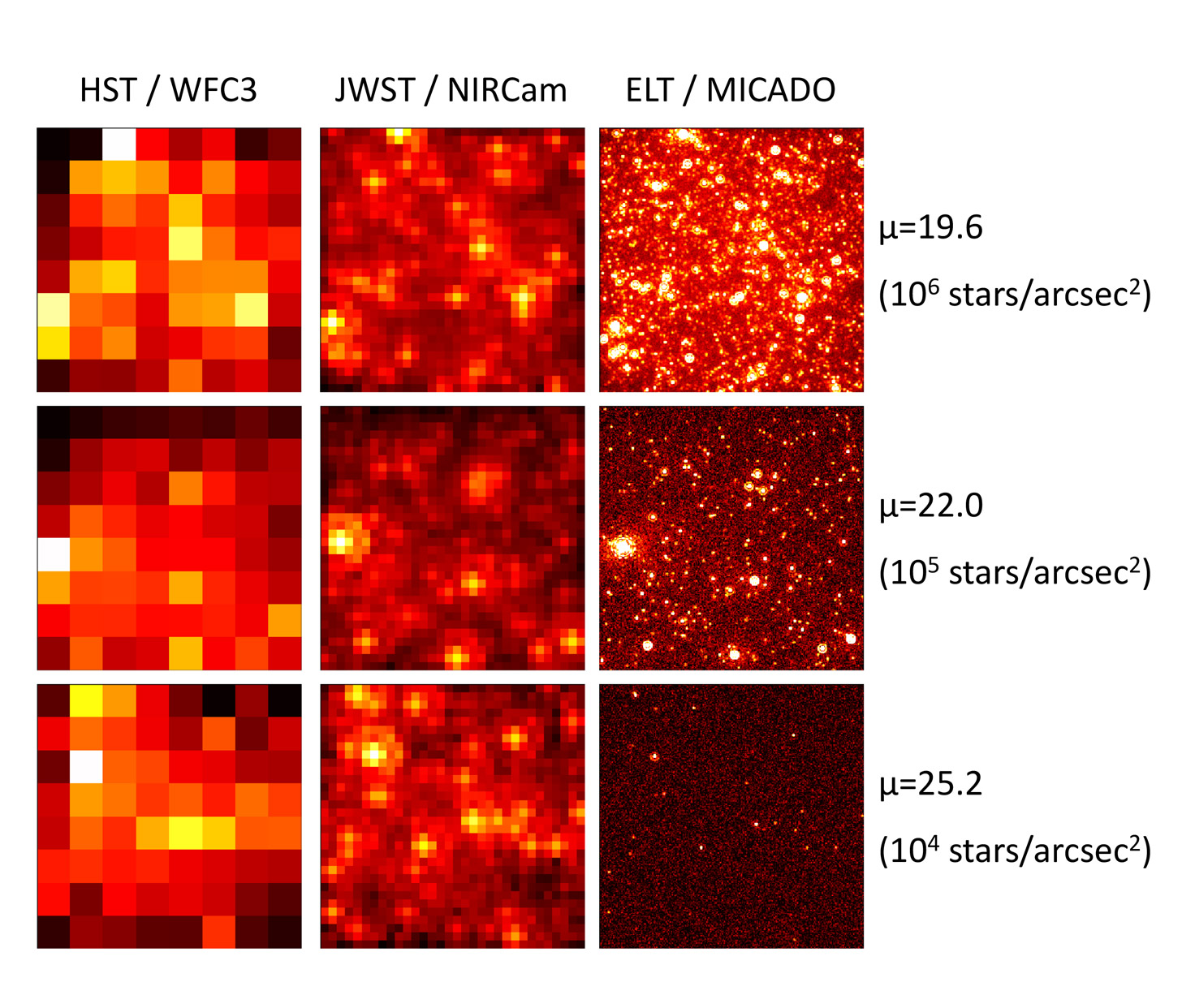}
\caption{Comparison of how crowded stellar fields would appear when observed by the HST (left), the JWST (centre), and MICADO (right). The bottom row matches the stellar density at 4-5 $R_{\rm eff}$ (the radius which encloses half the light) for NGC 4472 in the Virgo Cluster 
(i.e. at the edge of the galaxy) 
and represents the limit of JWST resolution. The top row corresponds to 2 $R_{\rm eff}$ in the same galaxy (i.e. 
closer to the centre) and many individual stars can still be measured by MICADO. Each panel is
1 arcsecond across. $\mu$ is the surface brightness in magnitudes per square arcsecond. These simulations were performed with ScopeSim. {\it Image credit: ESO/MICADO consortium (see also \cite{Davies2021}).}}
\label{fig:MICADO_sim}
\end{figure}

Stars with different characteristics including their evolutionary state,
chemical composition, and mass, emit nearly all the visible light in the Universe.
However, we can resolve and study thoroughly individual stars in stellar populations 
only in nearby systems. Similarly, only for these populations can we characterise 
properly their formation, evolution, and end products. 

With the ELT, astronomers will be able to study star-forming regions in unsurpassed 
detail at distances ten times 
larger than is possible today (Figure \ref{fig:MICADO_sim}). By enabling a closer look at how stars are born and 
evolve, the ELT will help us make substantial progress in the 
study of the early phases of star formation. Astronomers will use the ELT to measure 
the ratios of different chemical elements within stars, which will enable them to 
precisely measure their ages and chemical evolution. Comparing this against the stars' 
mass will help us better understand the lifecycle of different types of stars. With 
its extraordinary collecting power, the ELT will also be able to observe faint brown 
dwarfs, which trace the transition between stars and giant planets. 
The ELT will bring us closer to a full understanding of the evolution of planets and 
stars ranging from Jupiter mass to tens, and probably hundreds, of solar masses.  
Enormous stars end their lives violently in supernova explosions that spew out chemical 
elements that influence the formation of later stars and planets. Supernovae are some 
of the most luminous events in the Universe, meaning they can be seen at great distances 
and used as signposts of the Universe's evolution. The very biggest stars create 
$\gamma$-ray bursts when they explode; as the most energetic objects in the Universe 
these are also some of the most distant objects ever observed. The collecting power 
of the ELT will allow us to use them as distant lighthouses, shining through billions 
of years of evolution and taking us into the largely unknown early epoch of the 
Universe's Dark Ages. 

Using the ELT, for the first time, astronomers will be able to look beyond our Galactic 
neighbourhood to investigate individual stars in more distant galaxies and trace their 
history back to the very early Universe. As well as unravelling the complex formation 
and evolution of different types of galaxies, this will also allow us to compare our 
neighbourhood with other parts of the Universe.

\subsection{Black Holes}

\begin{figure}
\centering
\vspace{-3.8cm}
\includegraphics[width=0.7\textwidth]{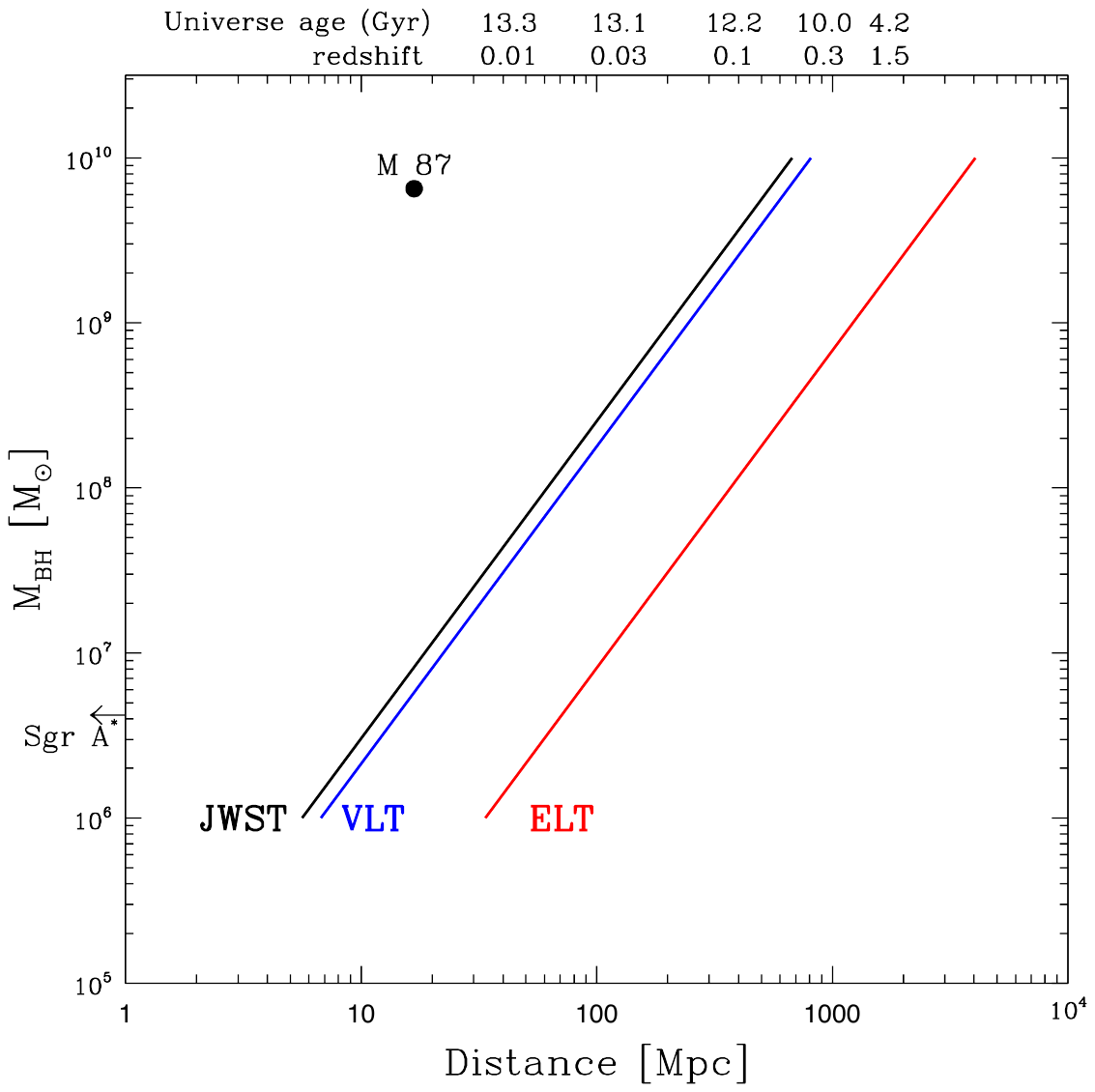}\hspace{5pt}
\vspace{-3cm}
\caption{Black hole mass vs. angular size distance. Spatial resolution 
assumes the diffraction limit case and scales with the telescope 
diameter. The ELT (red) line is for $\theta_{\rm res} = 10$ mas (a value 
intermediate between the two wavelength-dependent extremes), the 8 m - VLT (blue) one 
is for $\theta_{\rm res} = 50$ mas (ERIS), while the 6.5 m JWST (black) one is for 
$\theta_{\rm res} = 60$ mas. All lines correspond to the value required to resolve 
the so-called ``sphere of influence''. The regions of parameter space within reach of the
different facilities is the area to the left of the lines. For illustration, we show
the location of the SMBH in M 87, a giant elliptical galaxy in the Virgo constellation whose nucleus contains a black hole, the first ever to be directly imaged \cite{EHT_M87}.
Sgr A$^*$, the SMBH at the Galactic Centre
of the Milky Way, at a distance of $\sim 8$ kpc, is off-scale
to the left by two orders of magnitude. {\it Image credit: ESO.}} 
\label{fig:BH}
\end{figure}

Black holes are extremely common throughout the Universe. Indeed,
a supermassive black hole (SMBH) lies at the centre of almost every large galaxy, including the 
Milky Way, while some of the less massive black holes are thought to form when massive 
stars reach the end of their lives. Since these objects are black, i.e, they do not  
directly emit nor reflect light, astronomers rely mostly on indirect observations to 
spot their presence and study them. For example, they can infer much about a black 
hole by tracking the movements of stars and gas around it, something the ELT will 
excel at.   

The centre of the Milky Way is a unique laboratory for exploring gravity around the 
closest SMBH, Sagittarius A$^*$, 
with a mass $\sim 4 \times 10^6 M_{\odot}$. The ELT will enable astronomers to build on research 
done with ESO telescopes on the Galactic Centre, which was recognised with the 2020 
Nobel Prize in Physics\footnote{\url{https://www.eso.org/public/news/eso2017/}}.
A dense cluster of stars surrounds the SMBH, and the ELT will 
enable astronomers to study the behaviour of these stars in their strange 
environment with a level of detail and quality that we cannot reach 
with smaller telescopes. The Galactic Centre also provides us with a place to study 
the accretion of matter onto SMBHs, as well as to better 
understand the relationship between their activity and star formation at the centre 
of galaxies. 
ELT's MICADO will excel when compared with GRAVITY, the VLT interferometric instrument, 
which has already delivered several important results on the Galactic Centre
\citep[e.g.][]{GRAVITY2022}. 
While GRAVITY in fact offers a higher resolution, MICADO will have a much higher 
sensitivity and larger FoV, making it possible to track many more stars 
orbiting Sagittarius A$^*$ than currently possible. MICADO will deliver high-quality 
images with 12 mas resolution with a superb sensitivity over a FoV view tens of 
arcsecond across.

Another open 
question awaiting the advent of the ELT is the existence and demographics of intermediate 
mass ($100-10,000 M_{\odot}$) black holes. These black holes represent a link currently 
missing between stellar-mass black holes and SMBHs, and they could 
serve as seeds in the early Universe for the formation of the SMBHs 
that we see today. The ELT will be able to accurately measure the 3D velocities of stars 
in massive star clusters and dwarf galaxies, where these intermediate mass black holes 
are thought to reside, allowing astronomers to find out more about them.

The ELT will also study the role 
SMBHs play in the formation and evolution of galaxies and structures 
in the Universe. The centres of most galaxies harbour SMBHs with
masses $> 10^6 M_{\odot}$. SMBHs will be characterised out to large 
distances with the ELT, also through the resolution of the so-called 
``sphere of influence'', i.e. the region where black holes directly 
influence the motions of stars and gas through their gravity 
(Figure \ref{fig:BH}). For example, the volume within which the 
sphere of influence of a $10^9 M_{\odot}$ SMBH will be resolved will
increase by a factor of 75 as compared to what can be done now with the 
VLT, leading to a comparable increase in the number of accessible SMBHs.
This will allow us to trace the build-up of 
supermassive central objects in galaxies when the Universe was as young 
as a quarter of its present age. 

\begin{figure}
\centering
\includegraphics[width=0.99\textwidth]{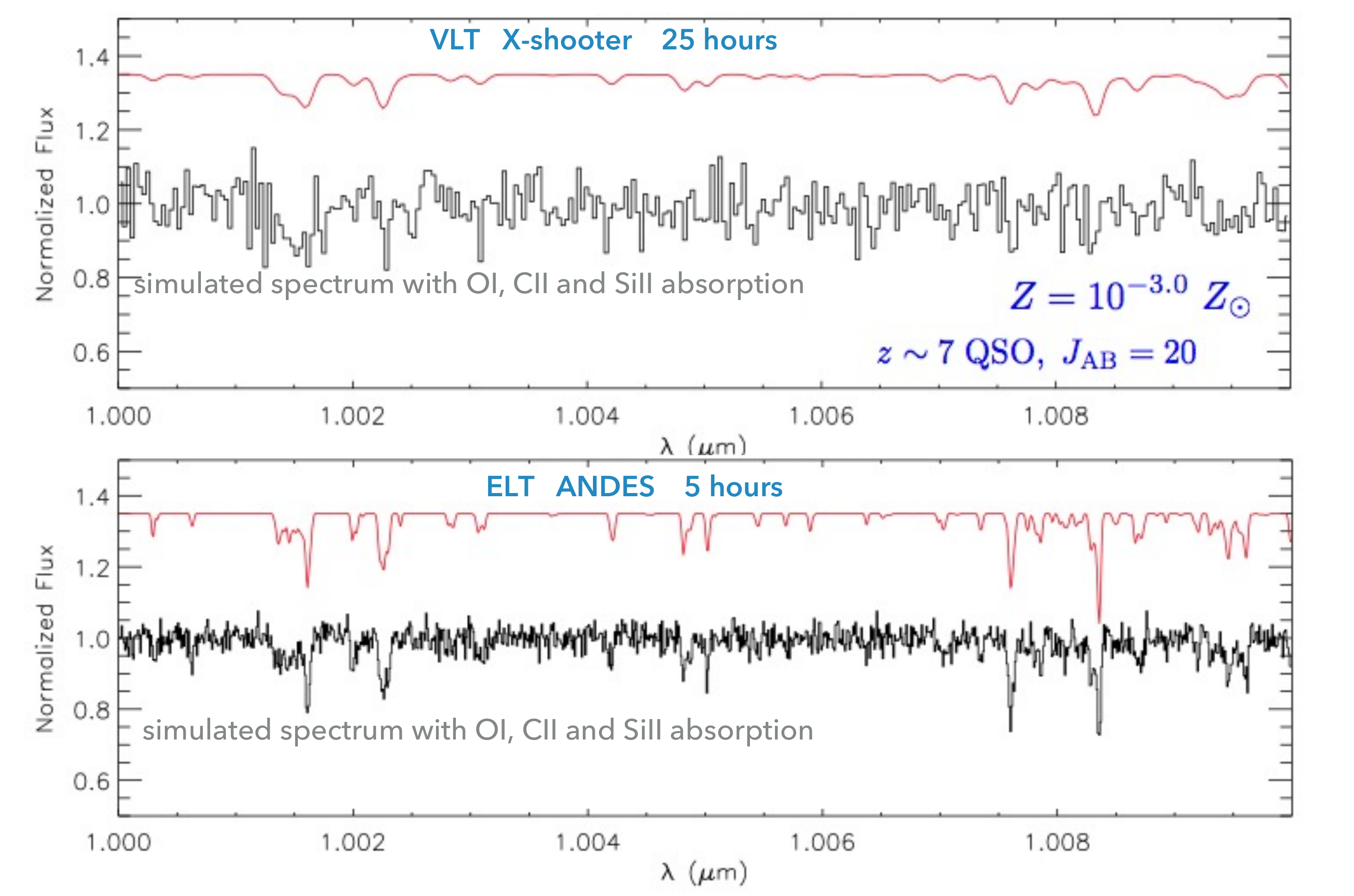}
\hspace{5pt}
\caption{Simulated observations of absorption due to O I, C II, and Si II at $z \sim 7$ using 8 m VLT+X-shooter (top) and ANDES (bottom). The thin red line at the top of each panel shows the predicted model spectrum generated from a cosmological simulation. The signals have been smoothed by the instrumental resolutions. Signal-to-noise ratios have been calculated assuming a 25 (5) hour integration with X-Shooter (ANDES) on a $z=7$ source with J$_{\rm AB} = 20$, similar to the quasar ULAS J1120+0641. No signal is detected in the X-shooter spectrum. In contrast, numerous narrow lines are detected with ANDES. {\it Image credit: Reproduced from ref. \cite{Maiolino_2013}, with permission.}}
\label{fig:ELT_XSHOOTER}
\end{figure}

\subsection{Galaxies}

\begin{figure}
\centering
\includegraphics[width=0.99\textwidth]{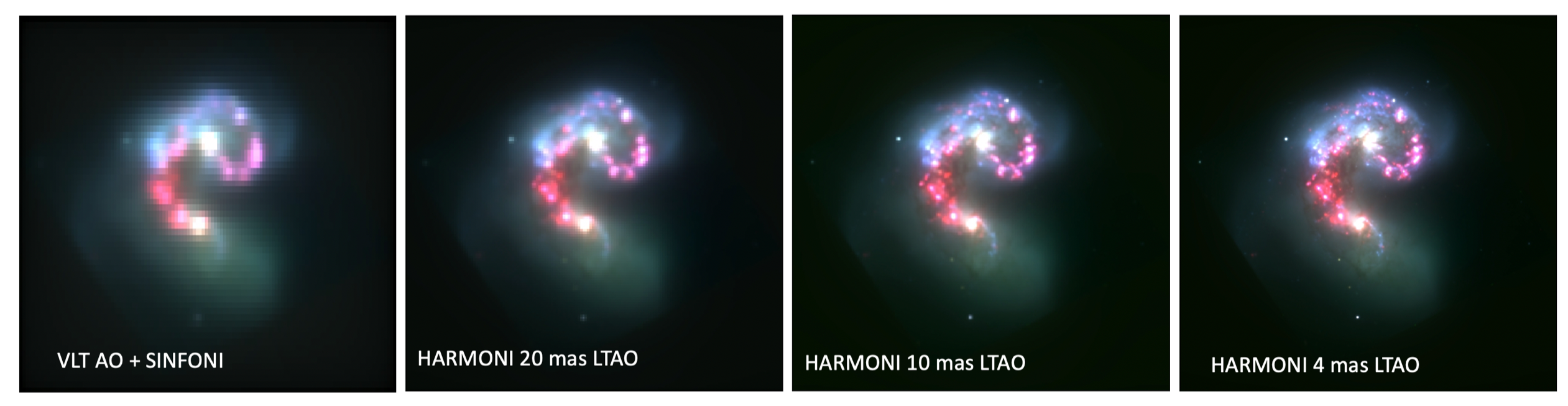}
\hspace{5pt}
\caption{A simulation of how a galaxy in the early universe would look like when observed with a VLT instrument and HARMONI
with three different spaxel scales and LTAO. Red is H$\alpha$, green is the V band, and blue is the B band. {\it Image credit: ESO/HARMONI consortium.}}
\label{fig:HARMONI_sim}
\end{figure}

Galaxy formation and evolution is driven by a number of mechanisms and a very complex interplay 
between them. These include the hierarchical merging of dark matter halos, the accretion and 
cooling of gas, and the gravitational fragmentation and the formation of molecular clouds. Star 
formation, nucleosynthesis, metal-enriched outflows driven by stellar winds, supernova 
explosions and energetic output from accretion onto SMBHs also influence 
galaxy formation and evolution. 

Thanks to the large number of recent deep surveys, it has become easier to derive galaxy 
properties such as luminosities, sizes, star formation rates, masses, and stellar masses,
as well as their redshift evolution. We now need more detailed spatially and spectrally resolved 
studies of individual galaxies to improve our understanding of their underlying physical 
processes. A comprehensive picture of galaxy formation should also describe a galaxy's internal 
structure, for example, how stellar populations and dust are distributed or what properties its 
central region has. Current facilities, however, are not able to describe these internal 
properties in detail in distant galaxies as they cannot reach the required small spatial scales,
i.e. that of giant molecular clouds; the ELT will enable us to do just this. 

HST has extended our view of evolving galaxies from the optical to the 
NIR, probing the emission from many stars in galaxies at high redshift and resolving 
the shapes and colours of galaxies. At the same time, IFUs, such as those 
at ESO's VLT, have routinely mapped the motion of ionised gas and the physical 
conditions of galaxies. The ELT will allow astronomers to observe fainter and 
more distant galaxies, up to when they were first building up their stellar 
populations and gas reservoirs, to improve substantially our current picture 
of galaxy evolution. The ELT will also study in great 
detail the gas in the intergalactic medium as revealed by the numerous 
absorption lines that are seen in the spectra of quasars 
(Figure \ref{fig:ELT_XSHOOTER}). 

Characterising galaxies with a wide range of masses, star formation activities, and 
surroundings requires the giant leap in spatial resolution uniquely provided by the ELT, 
with its 39 m main mirror and diffraction limit of 14 mas ($\sim$ 100pc at $z > 1$) in the 
K-band. Both imaging and IFU (with MICADO and HARMONI) will enable views of galaxies across 
cosmic times respectively $\sim$ 15 and $\sim$ 6 times sharper than possible with HST and 
JWST. HARMONI will overcome present limitations by tracing structures just a few tens of 
parsecs wide, and providing detailed studies by exploiting spectroscopic information with 
high angular and spectral resolution and sensitivity.
Figure \ref{fig:HARMONI_sim} shows a simulation of an
Antennae-like galaxy at $z \sim 2$ seen with a VLT instrument
and with HARMONI.

\subsection{Cosmology and Dark Matter}

Standard probes of the expansion of the Universe include weak gravitational lensing and the 
signature that light imprinted shortly after the Big Bang on today's distribution of galaxies, 
and type Ia supernovae. The ELT will contribute to current efforts to measure the acceleration of 
the Universe by characterising high-redshift Type Ia supernovae identified by JWST and other survey facilities. But the ELT will also map the expansion history of the 
Universe using a whole different method, namely by watching the Universe expand in real time. 

The current standard model of cosmology is the so-called Lambda-CDM model, 
wherein the Universe today is flat and consists of approximately 5\% baryons, 
$\sim 25$\% cold dark matter, and $\sim 70$\% dark energy. This implies that 
the Universe is not only expanding but also accelerating, due to the effects of 
a force never seen in the laboratory, which among other things works as an 
``anti-gravity'' by providing negative pressure. This requires entirely new 
physics, and consequently has sparked intense interest in mapping the expansion 
history of the Universe with high precision.

Extracting information about the Universe's expansion from standard probes relies on assumptions 
about the curvature of space, depends on the adopted cosmological model, and can only estimate 
the average expansion history over long time periods. A model-independent approach that measures 
the expansion rate directly was proposed as early as the 1960s \citep{Sandage1962}, but 
limitations in technology have meant that astronomers have not been able to make such a 
measurement in practice. This approach is known as redshift drift, and is a method that offers a 
truly independent and unique approach to exploring the expansion history of the Universe. The 
redshift of the spectra of distant objects is an indication of the expansion of the Universe, but
the change in this redshift over time is a measure of the change in the rate of 
expansion and depends on the underlying geometry: different Universes will 
display different drifts. For example, if the Lambda-CDM model is correct the 
drift will switch from positive to negative at $z \approx 2$ 
but if there is no dark energy the drift will always be negative \cite{Liske_2008}.

However, the estimated size of this redshift drift over a decade is only about 10 cm/s. 
Such a 
signal is about 10-20 times smaller than today's large telescopes can measure in such distant 
galaxies. The huge light-collecting area of the ELT, coupled with new developments in quantum 
optics to record ultra-stable spectra, means that this amazing measurement will lie within reach
of the ELT. All current methods to measure the expansion and shape of the Universe have to rely
on various intermediate steps, as in the case of type Ia supernovae.
Astronomers will use the redshift drift method with the ELT's ANDES instrument to determine the details of the accelerating expansion of the Universe and its shape directly and in real time, bypassing in one go all the 
intermediate steps, which are currently required to do this and thereby allowing us to quantify the nature of the dark energy responsible for the acceleration. 
Once a first epoch of observations is made, the redshift drift signal grows linearly with time. Hence, in the very long run (on the timescale of many decades) the redshift drift may well overtake the ability of other methods to constrain the expansion history of the Universe. 


Apart from the fundamental conceptual importance of directly observing the Universe's expansion, the redshift drift provides a new and crucial consistency test of the assumptions of our theories of cosmology. Two other such tests that will be significantly improved by the ELT are measurements of the temperature of the cosmic microwave background radiation and the primordial abundances of light elements. 

Astrophysical evidence for dark matter halos around galaxies first emerged in the 
1930s with studies of galaxy rotation curves, which plot the velocities of stars 
and gas against their distance from the centre of their host galaxy. Such studies 
still play an important role today. They require high-resolution observations of 
kinematics from the inner parts of galaxies to constrain the distribution 
of baryonic, or visible, mass. At higher redshifts, rotation curves are not 
currently resolved enough to constrain the fraction of dark matter. However, 
using data from the SINFONI and KMOS instruments on ESO's VLT, recent studies 
tried to do this in the inner regions of distant galaxies using both seeing 
limited and AO-assisted observations, as well as stacking several 
images together to reduce noise \citep{Genzel2020,Price2021}. Their results point 
towards a low fraction of dark matter, a finding that is supported by the 
observation of decreasing rotation curves. However, these results remain 
controversial because the signal-to-noise ratio is low, and the spatial 
resolution is rather coarse. 

ELT's HARMONI will be the first instrument able to reach the spatial resolution 
needed to disentangle the visible matter distribution from the dark matter 
distribution for high redshift galaxies. Mass model distributions will provide 
the measurements of dark halo central density and core radius for galaxies in a 
redshift range that is not reachable with 8-10 m class telescopes. 
Astronomers will be able to use the ELT to measure the shape of dark matter halos 
as a function of galaxy mass assembly history, cosmology, and environments. 
Detailed simulations show that HARMONI will be able to study spatially resolved 
kinematics with sufficient details and signal to noise ratio to perform mass 
models and recover the shape of dark matter halos down to stellar masses of 
$10^9 M_{\odot}$ at $z \sim 1.4$ and $10^{9.5} M_{\odot}$ at $z \sim 2.7$ 
using the H$\alpha$ line in just 2 hours of exposure. A multi-object spectrograph 
like the ELT's MOSAIC instrument will allow us to expand the statistical sample 
of galaxies, as well as extending studies of rotation curves and dark matter profiles.

\subsection{Fundamental Physics}

A fundamental constant of nature is any quantity whose value cannot be 
calculated using a given physical theory but can only be determined 
experimentally. Examples include the fine-structure constant $\alpha = 
e^2/(2 \epsilon_0 h c)$, where $e$ is the elementary charge, $\epsilon_0$ and $h$
are the electric and Planck constants, and $c$ is the speed of light 
(all of them also fundamental constants), which
quantifies the strength of the electromagnetic interaction between elementary 
charged particles, the proton-to-electron mass ratio, $\umu = m_{\rm p}/m_{\rm e}$, 
and the gravitational (or Newton's) constant $G$. Nothing prevents us from 
thinking that the laws of nature, which depend on these constants, may vary 
with time and location in the Universe. 

Astronomical observations can probe much longer timescales than Earth-based 
laboratory experiments and are therefore much more sensitive to possible time 
variations in the fundamental constants. By exploring the spectra of extremely 
bright distant quasars, for example, such possible variations can be probed 
over a large fraction of the history of the Universe \citep[see, e.g.][and 
references therein]{Murphy2022}. However, making these 
observations is very challenging, as it involves measuring the relative 
wavelength shifts of pairs of absorption lines whose wavelengths have 
different sensitivity to the fundamental constants, meaning that the wavelength 
calibration must be extremely accurate.

The ELT's ANDES ultra-stable high-resolution spectrograph will essentially remove 
the wavelength calibration uncertainties that plague current measurements and will 
vastly improve the constraints on the stability of fundamental constants. 
A discovery that the fundamental constants do vary across space and/or time would 
have huge implications. It would mean, for example, that Einstein's equivalence 
principle, which states that it is impossible to tell the difference between the 
effects of gravity and the effects of acceleration, because gravity is a 
fictitious force, due to the local geometry of spacetime, does not hold. 








\section{ELT Synergies}

Astronomy is currently experiencing a golden age, thanks to a combination of ground-based observatories and space telescopes that allow us to study the Universe in all electromagnetic bands, including also gravitational waves, neutrinos, and cosmic rays. The ELT 
will seamlessly integrate with a multitude of 
astronomical facilities covering the whole electromagnetic spectrum, 
which will include, besides the other two ELTs, ESO's VLTs, and ALMA,
the Square Kilometre Array 
(SKA)\footnote{\url{https://www.skatelescope.org/the-ska-project/}} 
in the radio, (hopefully still) JWST\footnote{\url{https://webb.nasa.gov/}} in the MIR, 
Euclid\footnote{\url{https://www.cosmos.esa.int/web/euclid}} (launched on July 1, 2023) in the visible and
NIR, the Vera C. Rubin Observatory (previously referred to as the Large Synoptic 
Survey Telescope [LSST])\footnote{\url{https://www.lsst.org/}} in the visible, 
PLATO\footnote{\url{https://www.cosmos.esa.int/web/plato}} in the visible, 
the Cherenkov Telescope Array 
(CTA)\footnote{\url{https://www.cta-observatory.org/}} in the
$\gamma$-rays, and many more. 

\section{Conclusions}

The ELT will be a revolutionary telescope and will be transforming Astronomy 
by excelling in collecting power and angular resolution, and by being an 
AO telescope and thereby providing space-like image sharpness. Located on Cerro Armazones 
in Chile, close to ESO's VLTs, it will also be the largest and likely the first of 
the ELTs, with its 39 m diameter and planned first light by the end of 2028. 

It will be an amazingly complex machine, with its huge dome, five mirrors, 
elaborate telescope systems and operations, and sophisticated instrumentation. 
It will have extremely exciting and unique scientific capabilities covering the 
whole of Astronomy, with a very strong community involvement and significant synergies
with many other astrophysical facilities. 
The ELT will open brand new windows on our Universe. Historically, such new windows 
caused revolutions in the way we understand Nature. We are ready for the next one. 

Please see the ELT Webpage (\url{https://elt.eso.org/}) for updated news and developments. 

\section*{Acknowledgement(s)}

Making the ELT a reality is only being possible thanks to a collaboration between a large 
number of different people in an enormous variety of roles, ranging from scientists and 
engineers to project managers and technicians, at ESO and across the ESO member states. The key 
responsibilities of some of the team members are shown at 
\url{https://elt.eso.org/about/team/}. This paper would not have 
been possible without them. There are many contractors and institutions working for the ELT
dealing with, e.g., the dome and the main
structure, the five mirrors, the instruments, the roads, 
software writing, consultancy, etc. A complete list can be found here: 
\url{https://elt.eso.org/about/industrial/}. 
We thank Richard Ellis and Olivier Hainaut for many useful comments, Michael Meyer for producing Figure \ref{fig:METIS}, and Alessandro Marconi and Valentina D'Odorico for
providing us with Figure \ref{fig:ELT_XSHOOTER}.

\section*{Disclosure statement}

No potential conflict of interest was reported by the author(s). 

\section*{Notes on contributor(s)}

{\bf Paolo Padovani}\\
\noindent
Paolo Padovani is Full Astronomer in the ELT Science office at ESO. 
After getting his PhD at the University of Padova, Italy, in 1989, 
he worked at the Space Telescope Science Institute (STScI), 
Baltimore, USA, at ESO, at the II University of Rome, Italy, 
and again at STScI for the European Space Agency.
He finally (re-)joined ESO in 2003. 
He has authored a number of ELT requirement documents and 
is currently the overall coordinator of the ELT Working Groups, which bring together expertise from within ESO, the Instrument Consortia, and the wider Community. 
He works on Active Galactic Nuclei over the whole electromagnetic
spectrum and has been recently expanding his research interests 
to Neutrino Astronomy.\\

\noindent
{\bf Michele Cirasuolo}\\
\noindent
Michele Cirasuolo is the ELT Programme Scientist at ESO  and is responsible for the scientific leadership of the telescope and its instrumentation. He has got a long experience with astronomical instrumentation and has been instrument scientist for the KMOS instrument and is principal investigator of the multi-object spectrograph MOONS for the VLT in Chile. With a PhD in astrophysics and two decades of active research, he has published extensively in the field of galaxy formation and evolution across cosmic time.



\bibliographystyle{tfnlm}
\bibliography{ELT}

\end{document}